\documentclass[twocolumn,showpacs]{revtex4}

\usepackage{graphicx}%
\usepackage{dcolumn}
\usepackage{amsmath}

\makeatletter
\def\btt#1{\texttt{\@backslashchar#1}}%
\DeclareRobustCommand\bblash{\btt{\@backslashchar}}%
\makeatother

\begin{document}

\preprint{MgB2 1/T1}

\title[Short Title]{Pressure-Temperature Phase Diagram of  Antiferromagnetism and 
Superconductivity in CeRhIn$_5$ and CeIn$_3$:\\$^{115}$In-NQR Study under Pressure}% Force line breaks with \\

\author{S.~Kawasaki$^1$, T.~Mito$^1$, G.~-q.~Zheng$^1$, C.~Thessieu$^{1,*}$, Y.~Kawasaki$^1$, 
K.~Ishida$^1$, Y.~Kitaoka$^1$, T.~Muramatsu$^1$, T.~C.~
Kobayashi$^2$, D.~Aoki$^3$, S.~Araki$^3$, Y.~Haga$^4$, R. Settai$^3$ and Y.~\=Onuki$^{3,4}$}

\affiliation{$^1$Department of Physical Science, Graduate School of Engineering Science, Osaka University, Toyonaka, Osaka 560-8531, Japan\\
$^2$Research Center for Materials Science at Extreme Condition, Osaka University, Toyonaka, Osaka 560-8531, Japan\\
$^3$Department of Physics, Graduate School of  Science, Osaka University, Toyonaka, Osaka 560-0043, Japan\\
$^4$Advanced Science Research Center, Japan Atomic Energy Research Institute, Tokai, Ibaraki 319-1195, Japan}

%%\date{today}% It is always \today, today, but you may specify any date with \date.

\begin{abstract}
We report the novel pressure($P$) - temperature($T$) phase diagram of antiferromagnetism and superconductivity in CeRhIn$_5$ and CeIn$_3$ revealed by the $^{115}$In nuclear-spin-lattice-relaxation ($T_1$) measurement.
In the itinerant magnet CeRhIn$_5$, we found that the N\'eel temperature $T_N$ is reduced at $P \geq$ 1.23 GPa with an emergent pseudogap  behavior.
In CeIn$_3$, the localized magnetic character is robust against the application of pressure up to $P \sim$ 1.9 GPa, beyond which the system evolves into an itinerant regime in which the resistive superconducting  phase emerges.
We discuss the relationship between the phase diagram and the magnetic fluctuations.

\end{abstract}

\pacs{PACS: 74.25.Ha, 74.62.Fj, 74.70.Tx, 75.30.Kz, 76.60.Gv} 
% PACS, the Physics and Astronomy Classification Scheme.
%\keywords{Suggested keywords}
                       
\maketitle
It has been reported that a superconducting (SC) order in cerium 
(Ce)-based heavy-fermion (HF) compounds takes place nearby the border at which an 
antiferromagnetic (AF) order is suppressed by applying pressure ($P$) to the 
HF-AF compounds CeCu$_2$Ge$_2$,\cite{CeCu2Ge2} CePd$_2$Si$_2$ \cite{Mathur} and CeIn$_3$ \cite{Grosche}. The superconductivity in these compounds, however, 
occurs only in extreme conditions where the  pressure exceeds $\sim$ 2 
GPa and temperature ($T$) is cooled down below $\sim$ 1 K. Indeed the 
experiments were restricted mainly to transport measurements. The discovery of 
$P$-induced HF superconductors in Ce-based HF-AF compounds has stimulated 
further experimental works under $P$ \cite{Hegger,Gegenwart,Kawasaki,Mito}. In 
order to gain profound insight into a relationship between magnetism and 
superconductivity in HF systems, systematic NMR/NQR experiments under $P$ 
are important, since they can probe the  evolution of the magnetic 
properties toward the onset of SC phase.

Recently, Hegger {\it et al.} found that a new HF material CeRhIn$_5$ consisting of alternating layers of CeIn$_3$ and RhIn$_2$ reveals an AF-to-SC transition at a relatively lower critical pressure $P_c$ = 1.63 GPa than in all previous examples \cite{CeCu2Ge2,Mathur,Grosche}.
The SC transition temperature $T_c$ = 2.2 K is the highest one to date among $P$-induced superconductors \cite{Hegger}.
This finding has opened a way to investigate the $P$-induced evolution of both magnetic and SC properties over a wide $P$ range.
In the previous paper \cite{Mito}, the $^{115}$In NQR study of CeRhIn$_5$ has clarified the $P$-induced anomalous magnetism and unconventional superconductivity.
In the AF region, the N\'eel temperature $T_N$ exhibits a moderate variation, while the internal field $H_{int}$ at $^{115}$In(1) site in the CeIn$_3$ plane due to the magnetic ordering is linearly reduced in  $P$ = 0 - 1.23 GPa, extrapolated to zero at $P^* = 1.6\pm 0.1$ GPa.
This $P^*$ is comparable to $P_c = 1.63$ GPa at which the SC signature appears \cite{Hegger}, which was indicative of a second-order like AF-to-SC transition rather than the first-order one suggested previously \cite{Hegger}.
At $P$ = 2.1 GPa, it was found that the nuclear spin-lattice relaxation rate $1/T_1$ reveals a $T^3$ dependence below the SC transition temperature $T_c$, which shows the existence of line-nodes in the gap function \cite{Mito}.
It is, however, not yet clear how the electronic states change with $P$ when the AF phase evolves into the SC phase.

On the other hand, CeIn$_3$ crystallizes in the cubic AuCu$_3$ structure and orders antiferromagnetically ($T_N$ = 10 K) at $P$ = 0 with an ordering vector {\bf Q} = (1/2,1/2,1/2) \cite{Morin}.
$T_N$ monotonically decreases with $P$ and superconductivity appears below $T_c \sim$ 0.2 K at a critical pressure $P_c$ = 2.55 GPa, but the onset of the superconductivity was observed {\it only} by the resistivity measurement and is limited in a narrow $P$ range of about 0.5 GPa \cite{Mathur,Grosche,Walker}.
The previous $^{115}$In-NQR result on CeIn$_3$ revealed that the AF order disappears around $P^*$ = 2.44 GPa close to $P_c$ and the Fermi-liquid behavior was observed below 10 K at $P$ = 2.74 GPa \cite{Kohori-2}.
However, no SC transition was seen.
Thus, the bulk nature of the $P$-induced SC state in CeIn$_3$ has not been confirmed yet by other measurements than resistivity.
This is in contrast with the case of CeRhIn$_5$ where the bulk nature of the SC phase was fully established \cite{Mito} which continues  up to 8 GPa \cite{Muramatsu}.

Here we report new $P$ - $T$ phase diagrams for CeRhIn$_5$ and CeIn$_3$ obtained through $^{115}$In-$T_1$ measurements and present a possible explanation for the different $P$ dependence of the SC phase in the two compounds.

The single crystals of CeRhIn$_5$ and CeIn$_3$ were grown by the self-flux and Czochralski Method, respectively \cite{Hegger,Onuki}, and were moderately crushed into grains in order to make rf pulses penetrate into samples easily. In a small piece of CeIn$_3$, the zero resistance transition is confirmed in $P$ = 2.2 - 2.8 GPa \cite{Muramatsu} (see below for detail).
At $P$ = 2.45 GPa, $T_c$ reaches a maximum of $T_c$ = 0.19 K and its transition width is the sharpest, which is in accordance with the previous results \cite{Mathur,Grosche}. The $T_1$ was measured at the transitions of 2 $\nu_{Q}$ ($\pm 3/2 \leftrightarrow \pm 5/2$) and 3 $\nu_{Q}$ ($\pm 5/2 \leftrightarrow \pm 7/2$) for $^{115}$In(1) site in CeRhIn$_5$ and of 3 $\nu_{Q}$ in CeIn$_3$, by the conventional saturation-recovery method.
The hydrostatic pressure was applied by utilizing a BeCu piston-cylinder cell, filled with Daphne oil (7373) as a pressure-transmitting medium.

Fig.1a shows some typical data sets for the $T$ dependence of 1/$T_1$ in CeRhIn$_5$ at $P$ = 0.46, 1.23 and  1.6 GPa.
In the high $T$ region, it is notable that $1/T_1$ becomes almost $T$-independent above the temperature marked as $T^*$. This indicates that Ce-$4f$  moment fluctuations are in a localized regime above $T^*$. In such a regime, $1/T_1$ is proportional to $p_{\rm eff}^2/J_{\rm ex}$ or $\sim p_{\rm eff}^2 W/J_{\rm cf}^2$ in terms of a localized moment picture. Here $p_{\rm eff}$ is an effective paramagnetic local moment, $J_{\rm ex}$ the RKKY exchange constant among $4f$ moments, $J_{\rm cf}$ the exchange constant between 4$f$ moments and conduction-electron spin, and $W$ is the bandwidth (BW) of conduction electrons.
A progressive suppression in $1/T_1$ with increasing $P$ is considered as due to the reduction in $p_{\rm eff}$ and/or the enhancement of $J_{\rm cf}$.
It is known in  HF systems that $T^*$ is scaled to the quasi-elastic linewidth in neutron-scattering spectrum, leading to a tentative estimation of the BW of HF state.
In CeRhIn$_5$, therefore, the incommensurate spiral order with a staggered moment of 0.267 $\mu_B$ occurs in an itinerant magnetic regime at $P$ = 0 \cite{Bao-2}.
As seen in Fig.1a, $T^*$ increases progressively with $T^*$ = 16, 23 and 27 K at $P$ = 0.46, 1.23 and 1.6 GPa, respectively. Since the application of $P$ further increases the hybridization between Ce-$4f$ state and conduction electrons, $J_{\rm cf}$, $J_{\rm ex}$, $T^*$ and BW are increased, whereas the size of the ordered moments is reduced.
In this case, one could expect that the slight increase in $T_N(P)\propto p_{\rm eff}^2(P)J_{\rm ex}(P)$ with increasing pressure is due to the enhancement of $J_{\rm ex}(P)$ which overcomes a possible reduction in $p_{\rm eff}$.

In order to focus on the itinerant magnetic regime in $T < T^*$, the $T$ dependence of $1/T_1T$ below 10 K is shown in Fig.1b for $P$ = 1.23 and 1.6 GPa. Above $T_N$, it is clearly seen that $1/T_1T$ shows a broad peak at $T_{PG}$. This resembles the pseudogap behavior found in the high-$T_c$ copper oxide superconductors \cite{Timusk}.
Namely, when $P$ approaches the critical pressures $P^*$ or $P_c$, the low-energy spectral weight of magnetic fluctuations is suppressed before an ordering occurs.
This is the first observation of the  pseudogap behavior in a magnetic region close to the SC phase.
We note that the pseudogap behavior has been found in either two- or lower-dimensional strongly correlated electron systems \cite{Timusk}.
Very recently,  anisotropic 3D AF fluctuation  was reported from neutron scattering with an energy scale of less than 1.7 meV for temperature as high as 3 $T_c$ \cite{Bao-2}. In fact, an anisotropic 3D character in the AF fluctuations has been revealed previously by NQR measurement in the related material CeIrIn$_5$ that is a superconductor at $P$ = 0 \cite{Zheng}. Thus the observation of the pseudogap behavior is consistent with the system bearing a magnetic  character of reduced dimensionality at low pressures. On the other hand, when the pressure is further increased, at $P$ = 2.1 GPa where the bulk SC transition appears, for example, $1/T_1T$ continues to increase down to $T_c$ = 2.2 K without any signature for the pseudogap behavior \cite{Mito}. The $T$ variation of $1/T_1T$ is consistent with the three dimensional (3D) AF Fermi-liquid model of the self-consistent renormalized (SCR) theory for nearly AF metals \cite{Moriya}.
Thus an evolution in the magnetic fluctuations, from a magnetic regime of reduced dimensionality to a more isotropic one, may take place in a narrow $P$ window of 1.2 - 2.1 GPa when the magnetic order evolves into the SC order. 

Next we deal with the results in CeIn$_3$. Fig.2a represents the $T$ dependence of $1/T_1$ in CeIn$_3$ at $P$ = 0, 1.42, 2.35 and 2.65 GPa.
It is seen clearly that the $T$ and $P$ dependencies of $1/T_1$ differ from  CeRhIn$_5$.
$1/T_1$ above $T_N$ is nearly $T$-independent at $P$ = 0.  At $P$ = 1.42 GPa, $1/T_1$ even increases upon cooling. Thus, the localized magnetic character in the paramagnetic state is robust against the application of $P$ in CeIn$_3$.
In the high pressure regime where the SC sets in, there seems to appear an itinerant regime below $T^* \sim$ 10 K.
$T^*$ eventually reaches $\sim$ 30 K at $P$ = 2.65 GPa.
Fig.2b indicates the $T$ dependence of $1/T_1T$.
For all pressures,  $1/T_1T$ increases down to $T_N$, giving no indication for the pseudogap behavior.
At $P$ = 2.35 GPa where the zero resistance is observed, it has  been confirmed, for the first time, that the magnetically ordering  survives with a relatively high value of $T_N$ = 5 K. The SC transition is, however, not found by the high-frequency ac susceptibility measurement down to 100 mK using the {\it in-situ} NQR coil, despite that a zero resistance was observed in the same sample at $T_c \sim$ 0.15 K.
From the present experiments, we conclude that the $P$ - $T$ phase diagram and the nature of the SC phase in CeIn$_3$ differs from in CeRhIn$_5$, in many aspects, reflecting their contrasting electronic and magnetic properties; we speculate that the SC phase in CeIn$_3$ that accompanies the Meissner diamagnetism, if any, is even narrower than suggested by the resistivity measurement.

The $P$ - $T$ phase diagrams for CeRhIn$_5$ and CeIn$_3$ are summarized in Fig.3.  For CeRhIn$_5$, $T^*$ slightly increases with increasing $P$ up to 1.0 GPa, as does $T_{N}$ which coincides with the previous result \cite{Hegger}.
However, $T_N$ decreases above $P$ = 1.23 GPa. At the same time, a pseudogap behavior emerges below $T_{PG} \sim 5.5$ K.
As $P$ is further increased, $T^*$ moderately increases with $d T^*/d P \sim$ 8 K/GPa. It is noted that $T^*$ is comparale to the temperature $T_{max,\rho}$ at which the resistivity exhibits a maximum value and also $d T^*/d P \sim d T_{max,\rho}/d P$ \cite{Hegger}.
We remark that $T^*$ and $d T^*/d P$ are also close to those in CeCu$_2$Si$_2$ \cite{Kawasaki} which is a superconductor at $P$ = $0$ and reveals a SC state over a wide $P$ region as well \cite{Thomas}.
In CeRhIn$_5$, the anisotropic 3D AF fluctuations may survive until the system is close to the SC state,   where the pseudogap behavior is emergent. 
Above $P^*$ or $P_c$ where the bulk superconductivity appears, the AF spin correlations  become more isotropic and the $T$ dependence of 1/$T_1T$ for $T > T_c$ can be explained on the basis of the 3D SCR theory for nearly AF metals \cite{Mito}.
While the isotropic AF fluctuation regime is fully established, the bulk SC is insensitive against $P$. In CeIn$_3$, by contrary, $T^*$ steeply increases above $P$ = 1.9 GPa, indicating an evolution of the system into an itinerant magnetic regime. It is interesting that the SC state becomes emergent in such a regime.
Close to $P_c$, the normal state resistivity, $\rho$ in CeIn$_3$ at low $T$ is also consistent with the 3D AF fluctuations model of itinerant magnetic regime \cite{Grosche}. However, when the pressure is further increased above $P_c$, a Fermi-liquid behavior of the resistivity returns more rapidly in CeIn$_3$ than in CeRhIn$_5$ and CeCu$_2$Si$_2$ \cite{Grosche}.
This is also corroborated by the observation of $T_1T$ = constant behavior at $P$ = 2.65 GPa in CeIn$_3$. Thus the window of the 3D AF fluctuation regime is much narrower in CeIn$_3$. We propose that the narrow SC region in CeIn$_3$ is due to the small window for the 3D AF fluctuation regime because of its large rate of $d T^*/d P$. This small window for the 3D AF fluctuation regime  against $P$ in CeIn$_3$ may be related to its cubic crystal structure which is more sensitive to external $P$ than a tetragonal structure.

Based on magnetically-mediated SC theoretical models \cite{Model}, it is predicted that 2D AF fluctuations are superior to 3D fluctuations in producing SC  \cite{Monthoux}. Therefore, the enhancement of $T_c$ in layered CeMIn$_5$ over CeIn$_3$ has been suggested to be due to their quasi-2D structure \cite{Hegger,Cornelius,Petrovic}. Our results suggest that the small window for the spin fluctuations regime in CeIn$_3$ may also be partly  responsible for its  lower $T_c$ .

In conclusion, on the basis of the $^{115}$In-NQR $T_1$ measurement, we have reported the $P$-induced evolution of the electronic and magnetic characteristics when approaching the SC phase in CeRhIn$_5$ and entering the SC phase in CeIn$_3$. In CeRhIn$_5$ that is already in the itinerant magnetic regime at $P$ = 0, $T_N$ slightly increases with $P$ at lower pressures which is in accordance with previous report \cite{Hegger}. However, $T_N$ starts to decrease
above $P$ = 1.23 GPa approaching the critical value at which SC sets in. At the same time, a pseudogap behavior emerges.
By contrast, in CeIn$_3$, the localized magnetic character is robust against the application of $P$ up to 1.9 GPa, beyond which $T^{*}$, which marks  an evolution into an itinerant magnetic regime, increases rapidly. It is interesting that the SC emerges in such an itinerant regime. The window for the 3D AF fluctuation regime with respect to external pressure is much narrower in CeIn$_3$ than in CeRhIn$_5$, which may also be partly responsible for its lower $T_c$ in CeIn$_3$.

We thank H. Harima for several useful conversations on this and related topics.
This work was supported by the COE Research (10CE2004) of Grant-in-Aid for Scientific Research from the Ministry of Education, Sport, Science and Culture of Japan. T.M., Y.Kawasaki and S. A. have been supported by JSPS Research Fellowships for Young Scientists.

$^*$Present address: Oxford Instruments, Tubney Woods, Abingdon, Oxon, OX13 5QX, England, UK.

\begin{figure}[htbp]
\caption[]{(a) $T$ dependence of $^{115}(1/T_1)$ in CeRhIn$_5$ at $P$ = 0.46, 1.23 and 1.6 GPa. The dotted lines are eye-guides. The dotted arrow indicates $T^*$.
 (b) $T$ dependence of $1/T_1T$.  In both (a) and (b), the solid and broken arrows indicate $T_{PG}$ and $T_N$ , respectively.}
\end{figure}
%fig2
\begin{figure}[htbp]
\caption[]{(a) $T$ dependence of $ ^{115}(1/T_1)$ in CeIn$_3$ at $P$ = 0, 1.42, 2.35 and 2.65 GPa. The dotted and broken lines are eye-guides. The dotted arrow indicates $T^*$.
(b) The $T$ dependence of $1/T_1T$. In both (a) and (b), the solid  arrows indicate $T_N$.}
\end{figure}
%fig3
\begin{figure}[htbp]
\caption[]{$P$ - $T$ phase diagrams (a) for CeRhIn$_5$ and (b) for CeIn$_3$.
(a) The open marks are determined from the resistivity measurements \cite{Muramatsu}, and the solid squares  are determined from the ac-$\chi$ measurement \cite{Mito}.
(b) The open marks for $T_N$ and $T_c$ are taken from the resistivity measurements\cite{Mathur,Grosche}.
The rest marks  are determined from the present work.
The inset indicates the detailed $P$ dependence of $T_c$ in expanded scales.
}
\end{figure}


\begin{references}
%1
\bibitem{CeCu2Ge2}
D.~Jaccard, K.~Behnia, and J.~Sierro, Phys.\ Lett.\ A {\bf 163}, 475 (1992).
%2
\bibitem{Mathur}
N.~D.~Mathur, F. M. Grosche, S. R. Julian, I. R. Walker, D. M. Freye, R. K. W. Haselwimmer and G. G. Lonzarich, Nature {\bf 394}, 39 (1998).
%3
\bibitem{Grosche}
F.~M.~Grosche {\it et al.}, Physica B {\bf 223-224}, 50 (1996); J. Phys. Condens. Matter {\bf 13}, 2845 (2001). J. Flouquet {\it et al.}, unpublished.
%4
\bibitem{Hegger}
H.~Hegger, C. Petrovic, E. G. Moshopoulou, M. F. Hundley, J. L. Sarrao, Z. Fisk and J. D. Thompson,  Phys. Rev. Lett. {\bf 84}, 4986 (2000).
%5
\bibitem{Gegenwart}
P. Gegenwart, C. Langhammer, C. Geibel, R. Helfrich, M. Lang, G. Sparn, F. Steglich, R. Horn, L. Donnevert, A. Link and  W. Assmus, Phys. Rev. Lett. {\bf 81}, 1501 (1998).
%6
\bibitem{Kawasaki}
Y. Kawasaki, K. Ishida, T. Mito, C. Thessieu, G.-q. Zheng, Y. Kitaoka, C. Geibel and F. Steglich, Phys. Rev. B{\bf 63}, R140501 (2001).
%7
\bibitem{Mito}
T.~Mito, S. Kawasaki, G.-q. Zheng, Y. Kawasaki, K. Ishida, Y. Kitaoka, D. Aoki, Y. Haga, and Y. \=Onuki, Phys. Rev. B{\bf 63}, 220507(R) (2001).
%8
%\bibitem{Kohori-1}
%Y.~Kohori, Y. Yamato, Y. Iwamoto, and T. Kohara, Eur. Phys. J. B {\bf 18}, 601 
%(2000).
%9
\bibitem{Morin}
P.~Morin, C. Vettier, J. Flouquet, M. Konczykowski, Y. Lassailly, J. -M. Mignot, and U. Welp, J. Low Temp. Physics {\bf 70}, 377 (1988).
%10
\bibitem{Walker}
I.~R.~Walker, F. M. Grosche, D. M. Freye, G. G. Lonzarich, Physica C {\bf 282-287}, 303 (1997).
%11
\bibitem{Kohori-2}
Y.~Kohori, T. Kohara, Y. Yamato, G. Tomka, P. C. Riedi, Physica B {\bf 281 \& 282}, 12 (2000).

%12
\bibitem{Muramatsu}
T.~Muramatsu {\it et al.}, submitted to J. Phys. Soc. Jpn (2001).
%16
\bibitem{Onuki}
Y. \=Onuki {\it et al.}, unpublished.
%17
\bibitem{Moriya}
T.~Moriya and T.~Takimoto, J. Phys. Soc. Jpn. {\bf 64}, 960 (1995).
%18
\bibitem{Cornelius}
A.~L.~Cornelius, A. J. Arko, J. L. Sarrao, M. F. Hundley, and Z. Fisk, Phys. Rev. B {\bf 62}, 14 181 (2000).
\bibitem{Bao}
Wei Bao, P. G. Pagliuso, J. L. Sarrao, J. D. Thompson, and Z. Fisk, Phys. Rev. B {\bf 62}, R14621 (2000).
\bibitem{Timusk}
For review see, T. Timusk and B. Statt, Rep. Prg. Phys. {\bf 62}, 61 (1999).
%19
\bibitem{Bao-2}
Wei Bao {\it et al.}, Cond-mat/0102503.
%14

%20
\bibitem{Zheng}
G.~-q.~Zheng, K. Tanabe, T. Mito, S. Kawasaki, Y.Kitaoka, D. Aoki, Y. Haga, and Y. \=Onuki, Phys. Rev. Lett. {\bf 86}, 4664 (2001).
%21
\bibitem{Thomas}
F.~Thomas {\it et al.}, Physica B {\bf 186-188}, 303 (1993).
%22
\bibitem{Model}
K. Miyake {\it et al.}, Phys. Rev. B {\bf34}, 6554 (1986); M.T. Beal-Monod,
 {\it et al.}, ibid. {\bf 34}, 7716 (1986); D. Scalapino {\it et al.}, ibid. 
{\bf 34}, 8196 (1986); P. Monthoux and D. Pines, {\bf 47}, 6069 (1993); T. 
Moriya and K. Ueda, J. Phys. Soc. Jpn. {\bf 63}, 1871 (1994).
%23
\bibitem{Monthoux}
P. Monthoux and G. G. Lonzarich, Phys. Rev. B {\bf 59}, 14598 (1999); ibid.
{\bf 63}, 054529 (2001).
%24
\bibitem{Petrovic}
C. Petrovic, R. Movshovich, M. Jaime, P.G. Pagliuso, M. F. Hundley, J. L. Sarrao, Z. Fisk, and J. D. Thompson, Europhys. Lett. {\bf 53}, 354 (2001).




\end{references}
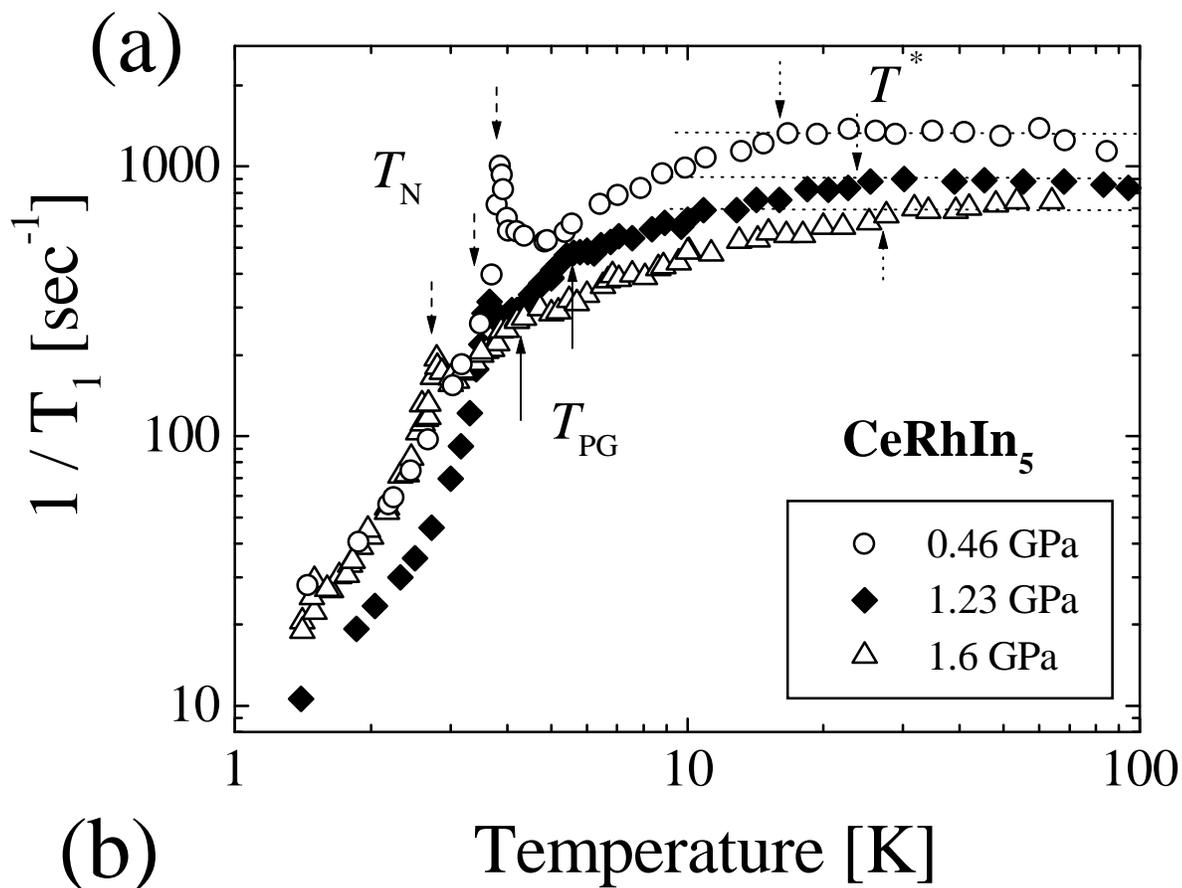
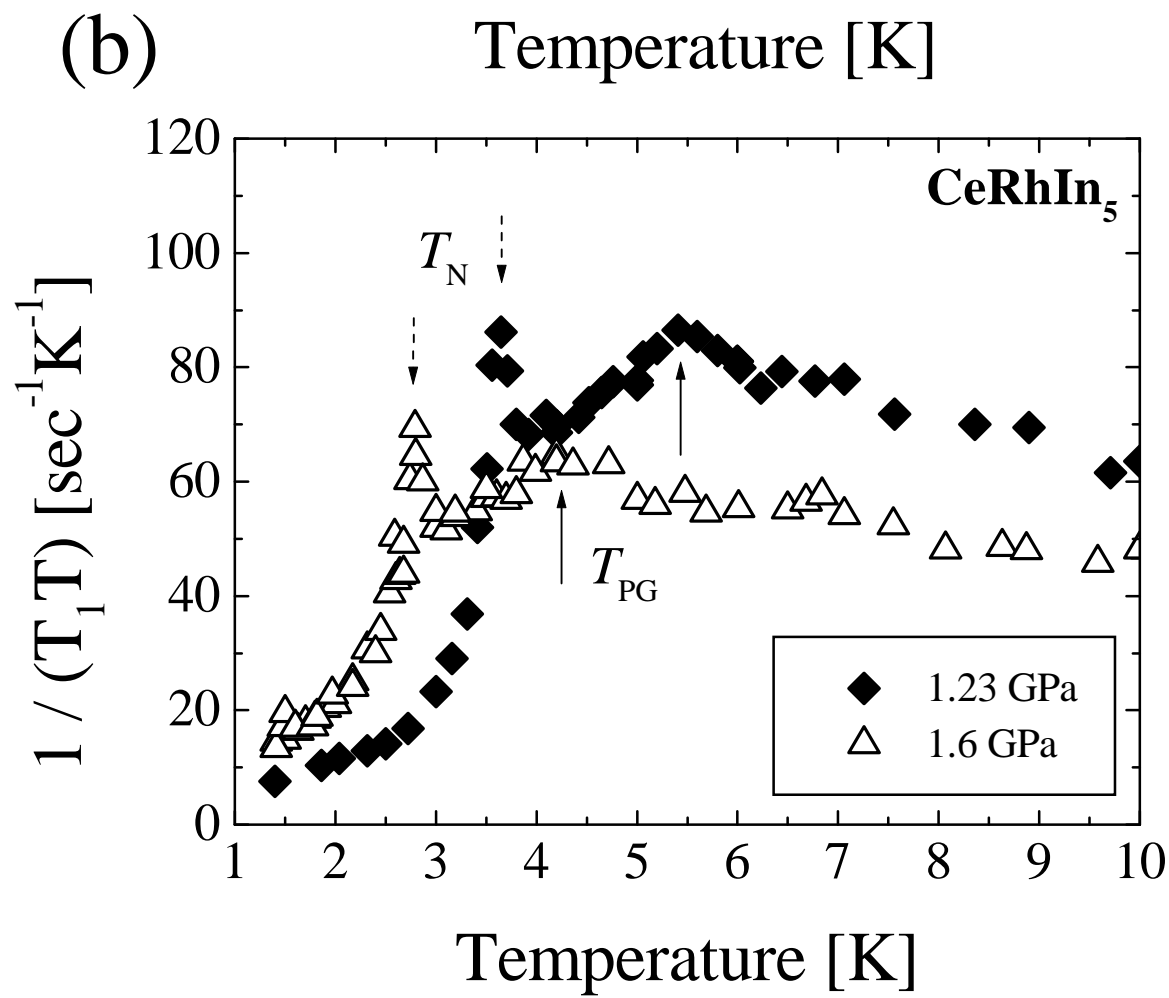
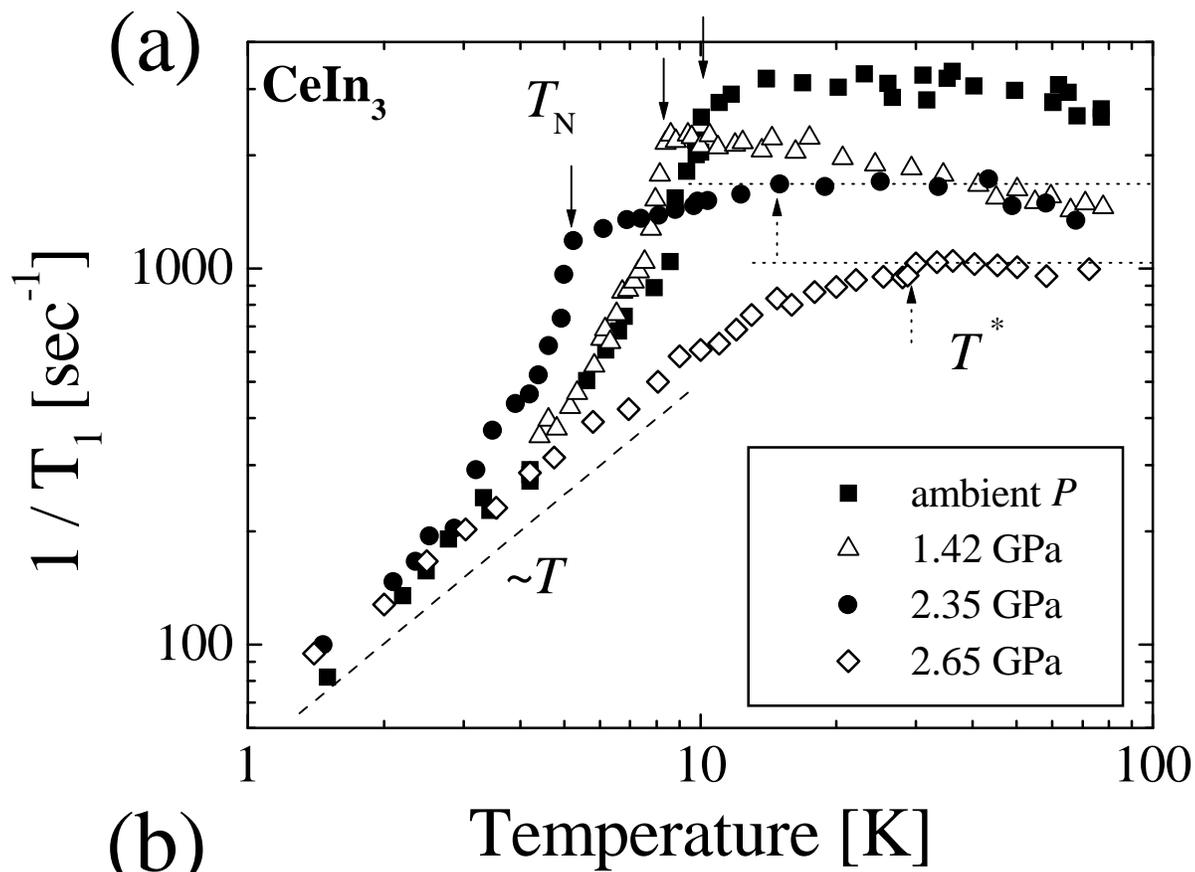
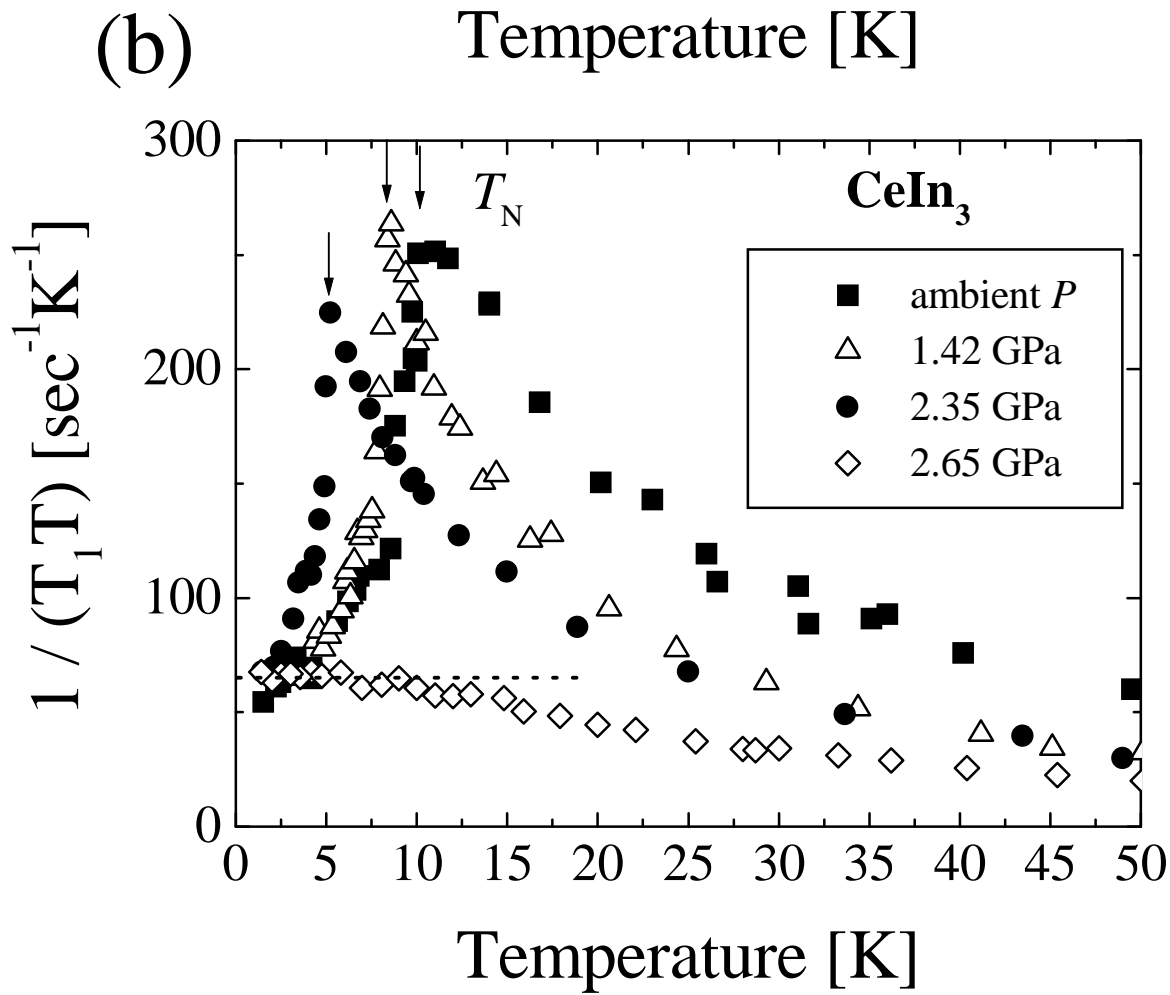
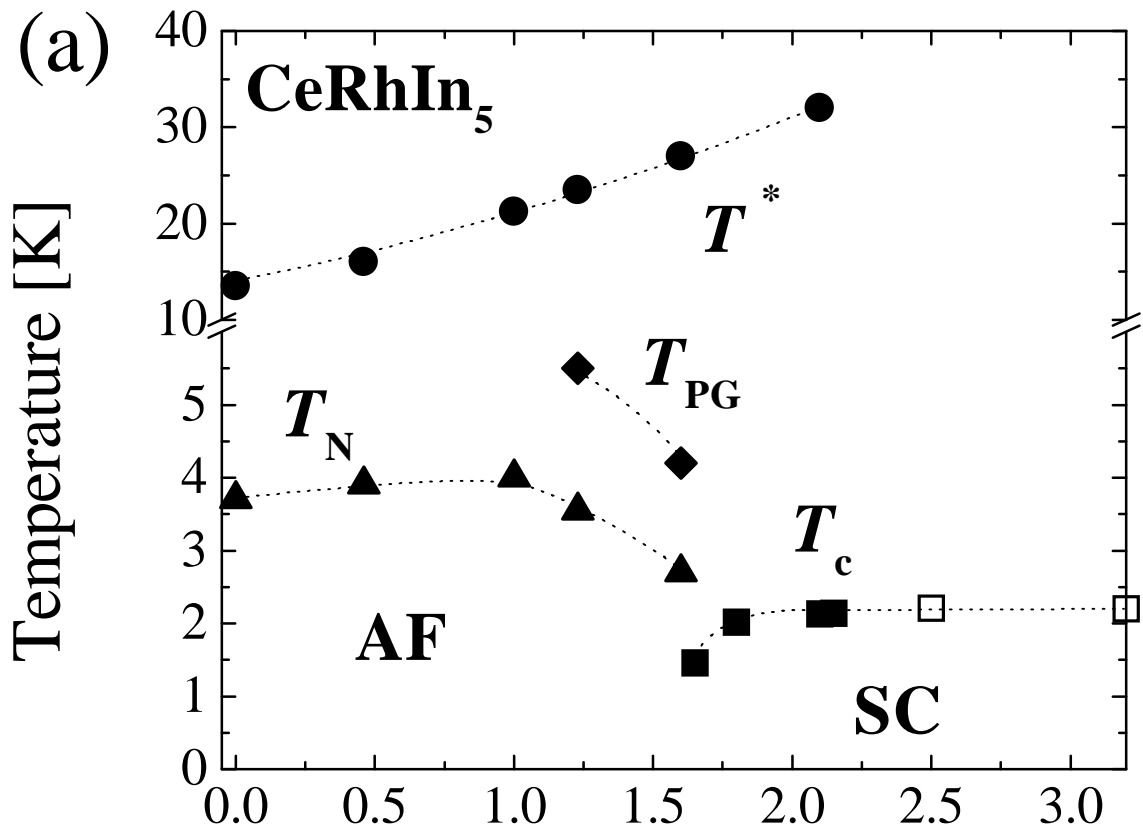
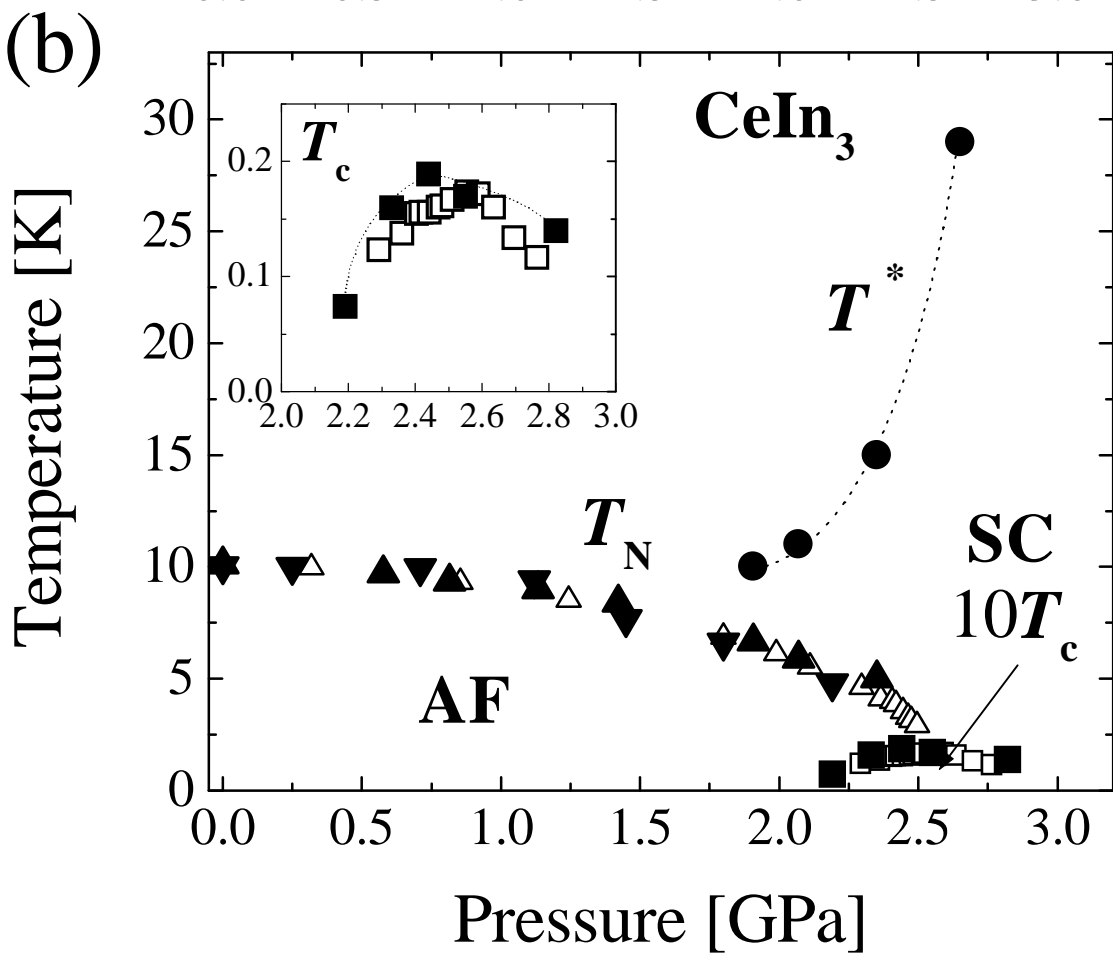
\end{document}